\documentclass[a4paper]{VisionStyle}
\usepackage{epsfig}
\begin{document}
\title{The XMM Large Scale Structure Survey:  \hspace{10cm}
Scientific Motivation and First Observations}
\author{M. Pierre\inst{1} \and I. Valtchanov\inst{1} \and
  A. Refregier\inst{2} }
\institute{Service d'Astrophysique, CEA/Saclay, F-91191 Gif sur
  Yvette, France; mpierre, ivaltchanov@cea.fr
\and
  Institute of Astronomy, Madingley Road, Cambridge CB3 OHA, UK;
  ar@ast.cam.ac.uk}
\maketitle
\begin{abstract}
Thanks to its unrivalled sensitivity and large field of view, XMM
potentially occupies a leading position as a survey instrument. We
present cosmological arguments in favour of a medium-sensitivity,
large-scale structure survey with XMM, using galaxy clusters as
tracers of the cosmic network. We show how this has motivated the
definition of a concrete survey, the XMM Large-Scale Structure
Survey (XMM-LSS), which will cover 64 square degrees with a
sensitivity about 1000 times better than the median ROSAT All-Sky
Survey. We present our predictions for cluster counts based on the
Press-Schechter formalism and detailed X-ray image simulations,
and show how they agree with the cluster statistics from recent
ROSAT cluster surveys. We also present the extensive
multi-wavelength follow-up associated with XMM-LSS, as well as the
first observations from the programme. 
\end{abstract}
\section{Introduction}
Ten years after the completion of the ROSAT All-Survey (RASS),
XMM, although not initially designed for it, is in a position to
open a new era for X-ray surveys. Its high sensitivity, good PSF
and large field of view, indeed makes XMM a unique instrument for
the study of extragalactic large scale structure (LSS).  In this
respect, two key points may be emphasized:

$\bullet$ Firstly, a high galactic latitude field observed with
XMM is ``clean'' as it contains only two types of objects, namely
QSOs (pointlike sources) and clusters (extended sources).

$\bullet$ Secondly, clusters more luminous  than $L_{[2-10]} 3
~10^{44}$ h$_{50}^{-2}$erg/s can be detected out to  $z = 2$ as
extended objects with a 10 ksec XMM exposure. With this exposure,
XMM   reaches a sensitivity of $\sim 10^{-14}$ erg/s/cm$^{2}$ in
[0.5-2] keV for extended sources (cf Fig. 4). This is about 1000
and 10 times deeper than the REFLEX  (\cite{boh}) and NEP
(\cite{hen1}) single area surveys respectively and provides a much
better angular resolution.

This makes XMM a powerful wide angle X-ray imager, with a
sensitivity which will remain unrivalled in the coming decade. In
particular, XMM is in a unique position to study  the evolution of
the cosmic network traced by clusters and QSOs. In Section 2, we
briefly review the cosmological implications of structure
formation models. In Section 3, we show how these translate into
the proposed XMM survey design. In Section 4, we discuss the
cosmological contraints which can be derived with the XMM-LSS. We
also compare our analytical predictions for the XMM-LSS cluster
counts with current observations of cluster statistics.  Section 5
describes the multi-wavelength follow-up associated with the XMM
survey. Finally, the first images from the AO-1 observations are
presented in Section 6.
\begin{figure*}[!ht]
\centerline{\psfig{file=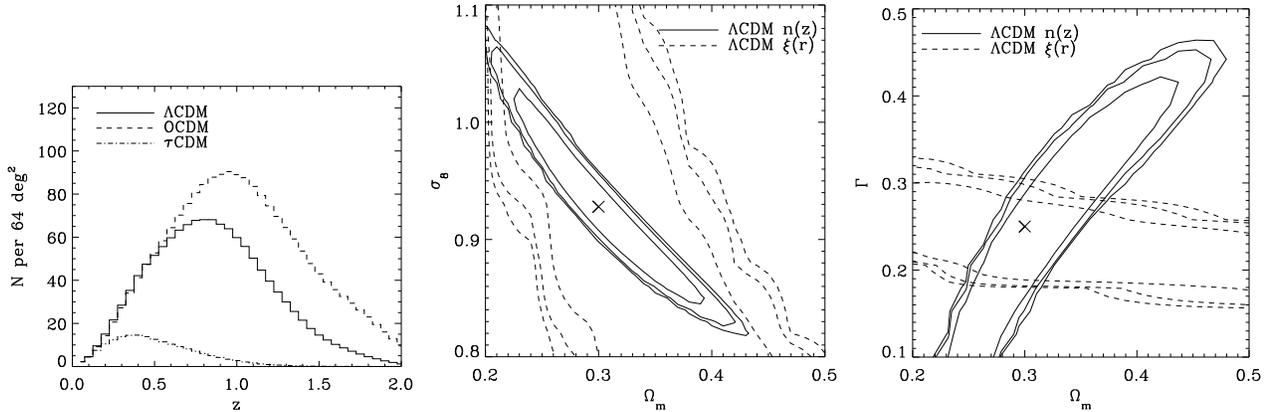,width=7cm,angle=90}}
\caption[]{{\bf Cosmological constraints from the XMM-LSS}(from
\cite{ref}). {\bf Left:} The predicted XMM-LSS cluster redshift
distribution for different cosmological models. The selection
function for the XMM-LSS, generated from image simulations, was
used to predict the observed counts. The currently favoured values
of the cosmological parameters were used for each model. Note
however that the resulting counts depend strongly on $\sigma_{8}$,
the normalisation of the matter fluctuations on 8 h$^{-1}$ Mpc
scale (see Fig. 7 of \cite{ref}); {\bf Centre:} Constraints upon
the cosmological parameters $\Omega_{m}$ and $\sigma_{8}$ for a
$\Lambda$CDM universe obtained from XMM-LSS cluster counts (solid
lines) and correlation function (dashed lines). In each case, the
68\%, 90\% and 95\% confidence level contours are shown along with
the assumed model (cross). Cluster abundance data provides strong
constraints upon the $\Omega_{m}-\sigma_{8}$ combination. {\bf
Right:} Constraints upon the cosmological parameters $\Gamma$ (the
shape of the matter power spectrum) and $\Omega_{m}$ for a
$\Lambda$CDM universe (symbols defined as before). The correlation
function is a powerful tool to constrain the shape of the initial
spectrum. These calculations have been performed assuming that
redshifts are available only over the $[0<z<1] \times [$64
deg$^{2}]$ volume. Note that the detection of a Coma-type cluster
within the XMM-LSS over the redshift range $1.5<z<2$ has a
probability of $\sim 6.5 \times 10^{-7}$ in the current
$\Lambda$CDM scenario.  Therefore, any such discovery in the
survey would put the currently favoured cosmological model in
severe observational difficulty. \label{mpierre-wb1_fig1} }
\end{figure*}

\section{Cosmological context}
Compared to galaxies, clusters of galaxies - the most massive
bound structures in the Universe - offer considerable advantages
for LSS studies, both because they can provide complete samples of
objects over a very large volume of space and because they are in
some respects simpler to understand. The haloes of clusters can be
traced by their X-ray emission (luminosity, size) while the theory
describing their formation (biasing) and evolution from the
initial fluctuations can be tested with numerical simulations. Such a
level of understanding does not exist for galaxies and even less so
for QSO formation. Studies of cluster LSS and cluster abundances
are thus powerful tools to constrain cosmological parameters,
independently of CMB and SNe studies. In addition, they can be used
to test the validity of fundamental assumptions of the standard
paradigm, such as the gravitational instability scenario.
Given the capabilities of XMM and the fact that the previous cluster
LSS survey (REFLEX, \cite{boh}) was limited to $z\leq 0.3$, we have
designed a project, the XMM-LSS, aiming at mapping the distribution of
matter out to redshifts $\sim 1-2$ and the evolution of the cosmic
network. The XMM observations constitute the core of the project, and
are complemented by an extensive multi-wavelength follow-up programme
to identify and study the astrophysical properties of the X-ray
sources in detail.

\section{Survey design}
\subsection{Defining the survey}
The following theoretical and practical constraints have led to the
current design of the XMM-LSS Survey :

$\bullet$ Detect a significant fraction of the cluster population
out to $z \sim 1$.

$\bullet$ Measure the cluster correlation function in two redshift
bins between $ 0 < z < 1$, with a good level of accuracy (about
15\% on the correlation length for both redshift intervals). This
implies a minimum of about 400 clusters for each bin. The
measurement of the evolution of the cluster correlation function
is a fundamental test of structure formation models, but has never
been done until now because of a lack of sufficiently wide and
deep surveys.

$\bullet$ Probe a comoving length which is significantly larger
than 100 h$^{-1}$Mpc at ~$ z \sim 1$, the characteristic scale in
the galaxy power spectrum of the local universe (e.g. \cite{lan}).
This constrain corresponds to an opening survey angle of $\sim
10^{o}$ at $z = 1$ (i.e. 400 h$^{-1}$Mpc)

$\bullet$ Find the best compromise between the  two above
constrains in order to minimise the necessary XMM observing time.

$\bullet$ Find an optimal survey location. An equatorial field is
optimal, as ground-based follow-up resources from both hemispheres
may be used. High galactic latitude and the absence of bright
X-ray sources (e.g. nearby clusters) are also required. Moreover,
the visibility of the field by XMM must be $\geq 15\%$. Given
this, only one area in the sky turned out to be favorable.

These different constraints led to the following lay-out: A
$8\times 8$ sq. deg. area paved with 10 ks XMM pointing separated
by 20 arcmin (i.e. 9 pointings per sq.deg.). The field is centered
around $\rm \alpha = 2^h20^m$, $\delta=-5^\circ$ (at
$b=-58^\circ$, with neutral hydrogen column $2 \times 10^{20} <
N_H/{\rm cm^{2}} < 5 \times 10^{20}$). The resulting sensitivity
is $\sim 3~10^{-15}$ erg/s/cm$^{-2}$ for pointlike sources in the
[0.5-2] keV band and of the order of $\sim 10^{-14}$ for extended
sources. This area surrounds two deep XMM surveys based on
guaranteed time: the XMM\_SSC/Subaru Deep Survey (80~ks exposures
in $1 \ \rm deg^2$) and the XMM Medium Deep survey (XMDS; 20~ks
exposures in $2 \ \rm deg^2$), the latter being a collaboration
between several instrumental teams: XMM-OM (Li\`ege), XMM-EPIC
(Milan-IFCTR), XMM-SSC (Saclay); MegaCam (Saclay, IAP); VIRMOS
(LAM, IFCTR, OAB). The area overlap will greatly assist in
quantifying the completeness of the survey.

\subsection{The XMM-LSS consortium}
The wide scope of the project has motivated the set-up of a large
consortium to facilitate both the data reduction/management and
the scientific analysis of the survey. The XMM-LSS Consortium
comprises the following institutes: Saclay (Principal
Investigator), Birmingham, Bristol, Copenhagen, Dublin,
ESO/Santiago, Leiden, Li\`ege, Marseille (LAM), Milan (AOB), Milan
(IFCTR), Munich (MPA), Munich (MPE), Paris(IAP), Santiago (PUC).
\section{Constraining cosmology}
\subsection{Analytical predictions}
We have studied quantitatively the prospects that the XMM-LSS cluster
catalogue offers for measuring cosmological parameters
(\cite{ref}). We used the Press-Schechter (PS) formalism to predict
the counts of clusters and their X-ray properties in several CDM
models. We computed the detection efficiency of clusters, using
realistic simulations of XMM images, and study how it differs from a
conventional flux limit. We computed the expected correlation function
of clusters using the extended halo model. Results are discussed in
Fig. 1 and can be summarized as follows:

$\bullet$ The cluster counts set strong constrains on the value of
the $\Omega_{m}-\sigma_{8}$ combination (the matter density and
the amplitude of mass fluctuations on 8 h$^{-1}$ Mpc scale). This
combination will also provide a consistency check for the
$\Lambda$CDM model, and a discrimination between this model and
the OCDM  model.

$\bullet$ The addition of the cluster 2-point correlation function
provides a constrain on $\Gamma$, the shape of the initial density
fluctuation power spectrum.

$\bullet$ With the current survey design, the $simultaneous$
expected precision on  $\Omega_{m}, \sigma_{8}$ and $\Gamma$ is
about 15\%, 10\%, 35\% respectively.

While these predictions are based on analytical calculations and
image simulations, it is worth emphasizing that they agree with
the latest measurements of XMM specifications and cluster
statistics. Firstly, XMM has now proved to reach its nominal
sensitivity in normal observing conditions. Secondly, we have used
in our predictions, a value of $\sigma_{8}=0.93 \pm 0.07$ for
$\Omega_m = 0.3$, as measured with the cluster temperature
function (\cite{eke}). For the currently favoured $\Lambda$CDM
model, this yields some 370 and 520 clusters detected in the
$0<z<0.6,~0.6<z<1$ intervals respectively, having a temperature
greater than 2 keV.
The present uncertainties on $\sigma_{8}$ globally result in a
factor of 2 on our predicted cluster numbers (i.e. from 600 to
1200 clusters detections expected within $0 < z < 1$). The high
sensitivity to $\sigma_{8}$ is not surprising, as it is precisely
that which makes cluster counts a good measure of this parameter.
This uncertainty can be reduced by analysing about 10 sq.deg. of
the XMM-LSS, the minimum area required to improve upon the current
measurements of $\sigma_{8}$ in the presence of shot noise and
cosmic variance. This has practical consequences for the number of
optical spectroscopic nights needed to measure cluster redshifts.
However, beyond the current nominal value of $\sim$ 15 clusters
per sq.deg, a random sampling of the cluster population would be
sufficient to fulfill our cosmological goals.\\
\subsection{Comparison with current observations}
To further validate our predictions, we have compared our
analytical model with the current measurements of cluster
statistics. For this purpose, we first considered the cluster
luminosity function $F(L)$. Figures 2 and 3 show the measurement
of $F(L)$ from the ROSAT Deep Cluster Survey (RDCS, \cite{ros}) at
$z=0$ and $z=0.8$. As can be seen on the figures, the measured
luminosity function was shown not to evolve in this redshift
interval. Note that, in Figure 3, the RDCS $F(L)$ was extrapolated
to low luminosities ($L_{[0.5-2]} \simeq 1.5 ~10^{43}$
h$^{-2}_{50}$erg/s (EdS) at $z = 0.8$) in order to match the
XMM-LSS flux limit ($\sim 10^{-14}$ erg/s in [0.5-2] keV) which is
about 3 times lower than that of RDCS ($L_{[0.5-2]} \simeq 4
~10^{43}$ h$^{-2}_{50}$erg/s (EdS) at $z = 0.8$, \cite{ros}). We
also calculated the analytical luminosity function using the mass
function from the Press-Schechter formalism for haloes with $T> 1
keV$, the $M-T$ relationship assuming virial equilibrium, and the
$L-T$ relationship measured by \cite{arn}. As can be seen on
Figures 2 and 3, the comparison with the observed luminosity
function shows good agreement. Indeed, the agreement is better
than a factor of 2 for bolometric luminosities lower than
$10^{44}$ erg/s, which represent the bulk of our cluster
population. As shown in Figures 2 and 3, the Sheth \& Tormen
(1999) formalism, which provides a better fit to N-body
simulations than PS, further improves the match for high
luminosity objects.
\begin{figure}[ht]
\centerline{\psfig{file=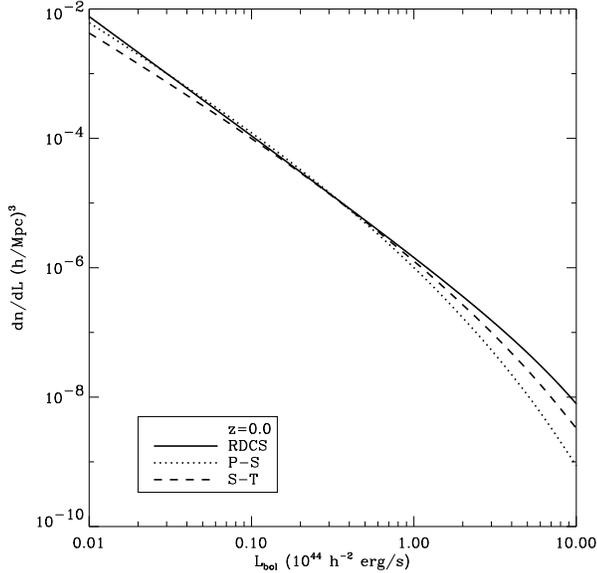,width=8cm,angle=0}}
\caption[]{ The cluster luminosity function at $z = 0$. Solid
line: observed from RDCS. Dotted line: our predictions using the
PS formalism. Dashed line: same, but using Sheth-Tormen formalism
(\cite{she}), which provides a better fit to N-body simulations.
 \label{mpierre-wb1_fig2}}
\end{figure}
\begin{figure}[ht]
\centerline{\psfig{file=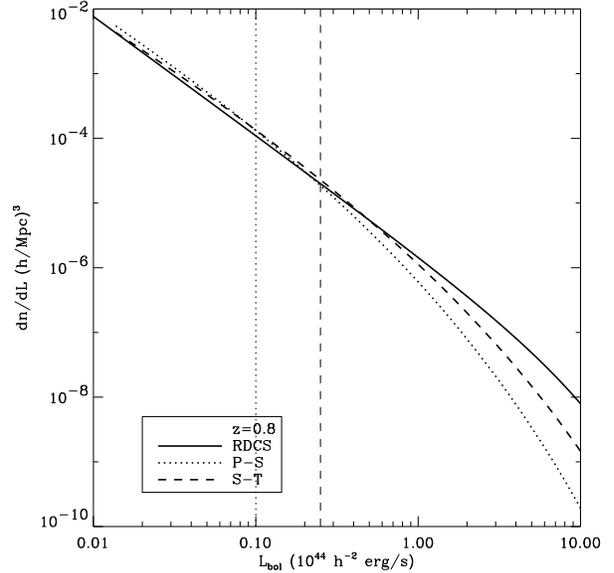,width=8cm,angle=0}}
\caption[]{Same as Fig. 2, but for  $z = 0.8$. Here the RDCS
$F(L)$ has been extrapolated beyond $L_{bol}\simeq 2.5 ~10^{43}$
erg/s ($\Lambda$CDM), which corresponds to $ L_{[0.5-2]}\simeq 4~
10^{43}$ h$^{-2}_{50}$erg/s (EdS), the current limit of the RDCS
at $z=0.8$ (vertical dashed line). The vertical dotted line
represents the XMM-LSS detection 90\% limit for clusters at $ z =
0.8$. \label{mpierre-wb1_fig3} }
\end{figure}

As a more practical test, we also compared our predicted cluster
counts $N(z)$ with that obtained by a direct integration of the
observed luminosity function. For the latter, we adopted the same
flux limit as above (using redshifted Raymond-Smith spectra), and
used the non evolving $L-T$ relationship from \cite{arn}. We also
extrapolated $F(L)$ at high redshifts to lower luminosities (as
described above) and extended the validity of the RDCS $F(L)$ from
$z = 0.8$ to $z = 1$ (the minimum luminosity corresponding to our
flux limit is $L_{[0.5-2]} \simeq 3 ~10^{43}$ h$^{-2}_{50}$erg/s
at $z = 1$). As shown in Figure 4, the two distributions are in
good agreement. In the present calculation, we assumed that RDCS
is not evolving (\cite{ros}); there is however growing evidence
that massive clusters are less numerous at high redshift
(\cite{hen2}). This would improve further the agreement with our
simulations. Consequently, our analytical model is consistent with
the current measurements of cluster statistics, and thus provides
secure predictions for the cluster counts in the XMM-LSS.

\begin{figure}[ht]
\centerline{\psfig{file=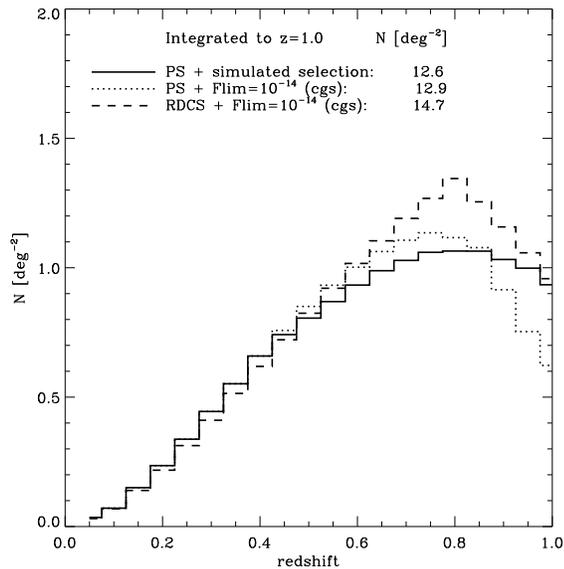,width=8cm,angle=0}}
\caption[]{ Comparison of analytical predictions for the cluster
number count to that derived from a direct integration of the
observed luminosity function. Solid line : \cite{ref} predictions
using PS formalism and detailed image simulations. Dotted line :
same, but using a flux limit of $10^{-14}$ erg/s/cm${2}$ in
[0.5-2] keV. Dashed line: predicted $N(z)$ using the RDCS
luminosity function $F(L)$ extrapolated down to $L_{0.5-2} \simeq
1.6 ~10^{43}$ h$^{-2}_{50}$erg/s (EdS) in order to reach the
XMM-LSS flux limit of $\sim 10^{-14}$ erg/s/cm$^{2}$ out to $z =
0.8 $; also $F(L)$ is assumed not to evolve   out to $z=1$. At $z$
close to 1, \cite{ref} predict less clusters than expected from
$F(L)$ alone, since they take evolution into account. $\Lambda$CDM
cosmology was assumed in all cases. \label{mpierre-wb1_fig4}}
\end{figure}

\section{Multi-wavelength follow-up}
\underline{Imaging:} The imaging of the $8 \times 8 \ \rm deg^2$
XMM-LSS area is the priority target of the Canada-France-Hawaii Legacy
Survey\footnote{http://cdsweb.u-strasbg.fr:2001/Instruments/Imaging/\\
Megacam/MSWG/forum.html}. MegaCam is a one degree field imager built
by CEA to be installed at the new CFHT prime focus. It will come into
operation by mid-2002. It will provide the deep high quality optical
multi-color imaging counterpart of the X-ray sources (u*=25.5,
g'=26.8, r'=26.0, i'=25.3, z'=24.3) at a rate of 15 deg$^{2}$/yr in at
least three colours. In particular, an optical cluster catalogue is
currently under construction using the CFH12K (and later MegaCam)
observations using both spatial clustering analysis and multi-color
matched filter techniques, in addition to photometric redshift
estimates. Moreover, the MegaCam data will form the basis of a weak
lensing analysis\footnote{
http://www.iap.fr/LaboEtActivites/ThemesRecherche/\\
Lentilles/LentillesTop.html}, whose cosmological constraints will be
compared to that provided by the X-ray data on the same region. This
will be the first, coherent study of LSS on such scales. R and z'
imaging from CTIO are also being analzed. Data pipelines and
processing have been developed by the
TERAPIX\footnote{http://terapix.iap.fr} consortium; this will provide
object catalogues and astrometric positions for the entire surveyed
region. In addition, deep NIR VLT imaging (J, H, K ) of $1<z<2$
cluster candidates found in in the XMM-LSS will be performed as a
confirmation prior to spectroscopy.

\underline{Spectroscopy:} The standard spectroscopic follow-up is
designed to perform redshift measurements for all identified
$0<z<1$ X-ray clusters in Multi-Object-Spectroscopy mode, using
the 4m and 8m telescopes to which the consortium has access. We
plan to take 1 mask per cluster, randomly sampling the XMM AGN
population at the same time, as well as the surrounding
filamentary galaxy distribution connecting clusters. This mapping
around $0<z<1$ clusters will have an important scientific
potential for studies of galaxy environments and bias. We shall
subsequently undertake programmes of advanced spectroscopy that
will focus on individual objects, and include high resolution
spectroscopy, the measurement of cluster velocity dispersions, QSO
absorption line surveys, as well as NIR spectroscopy of our $z>1$
cluster candidates.

\underline{Radio:} In the radio waveband, the complete survey region
is being mapped using the VLA at 74MHz and 325MHz. Radio coverage is
not only particularly relevant for tracing merger events triggered by
structure formation, but also as a useful indicator of galactic
nuclear or star--formation activity.

\underline{Sunyaev-Zel'dovich:} observations (S-Z) are also
planned. Clusters in the XMM-LSS field will be targets of the
prototype OCRA (One-Centimeter Radiometer Array) instrument from
2002. The full XMM-LSS field will be mapped by the complete OCRA, and
will be an early target of the Array for Microwave Background
Anisotropy (AMiBA) after 2004 (\cite{lia}).  This will enable a
statistical analysis of the physics of the ICM as a function of
redshift. In the long term, these observations will also provide
invaluable information on the low density structures such as cluster
outskirts and groups, and their connections to supercluster
filaments. These measurements are complementary to the X-ray and weak
lensing surveys, connecting the mass distribution of clusters to the
structure of the hot gas they contain. The three data sets together
will also provide a direct and independent check of the extragalactic
distance scale.

\underline{Infrared}: In the infrared, the SWIRE\footnote{
http://www.ipac.caltech.edu/SWIRE} SIRTF Legacy Programme will cover
10 deg$^{2}$ of the XMM-LSS in 7 wavebands from 4 to 160 mm. The
estimated IR source numbers in this area are around 20000/900/250 and
700/50/500 for starbursts/spiral-irregular/AGN in the $0<z<1$ and
$1<z<2$ redshift intervals, respectively. This represents a unique
X-ray/IR combination in depth and scales to be probed. The coordinated
SWIRE/XMM-LSS observations will clarify an important aspect of
environmental studies, namely how star formation in cluster galaxies
depends on the distance to the cluster centre, on the strength of the
gravitational potential, and on the density of the ICM. In this
respect the XMM-LSS represents the optimum SWIRE field, where galaxy
environment, deep NIR imaging and optical spectroscopic properties
will be the main parameters in modelling the MIR/FIR activity.  Here
also, the location of IR AGNs within the cosmic web will help
establish their nature. The FIR/X/optical/radio association will also
provide unique insights into the physics of heavily obscured objects,
as well as the first coherent study of biasing mechanisms as a
function of scale and cosmic time, for X-ray hot (XMM), dark (weak
lensing), luminous galactic (optical/NIR) and obscured (SWIRE)
material.

In summary, the XMM-LSS multi-$\lambda$ data set will offer the first
evolving view of structure formation from supercluster to galaxy
scales. Its comprehensive approach constitutes a decisive new step in
the synergy between space and ground-based observatory resources and
therefore a building block of the forthcoming Virtual Observatory.
\section{First XMM observations}
For the XMM AO-1, about 700 ks have been allocated to the project
(GT and GO time) corresponding to some 50 pointings. First
pointings were performed in July 2001. Some of them are presented
on Fig. 5. Our first analysis of the data confirms our predictions
for the XMM sensitivity with 20 ks. Fig. 6 shows two examples of
extended sources found in the preliminary analysis: one compact
group and one distant cluster candidate. The overlays give a good
impression of the mean source density and emphasize the need for
multi-color ground-based imaging in the process of source
identification.
\begin{figure*}[ht]
\centerline{\psfig{file=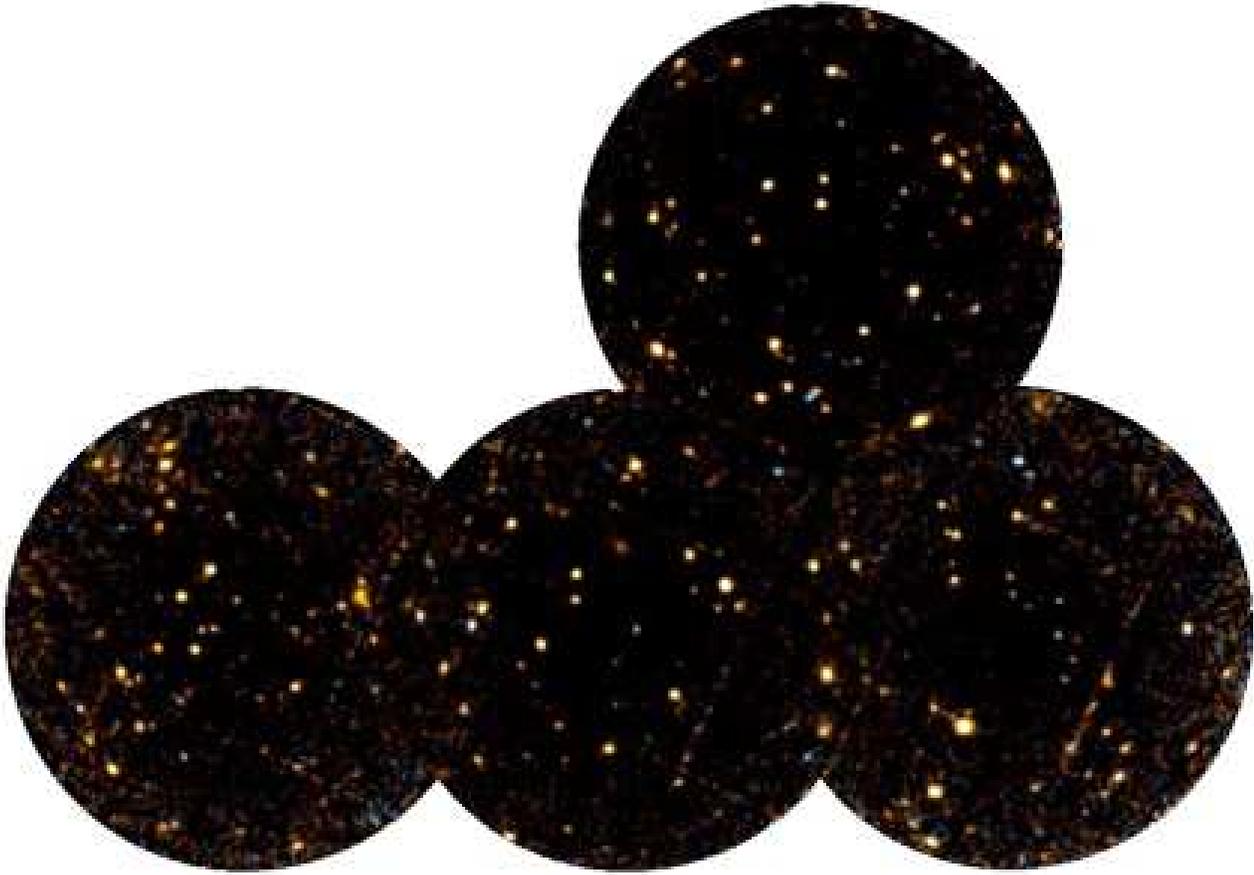,width=18cm,angle=0}}
\caption[]{First XMM pointings of the survey obtained during the
Guaranteed Time part of the programme led by the
Li\`ege/Milano/Saclay groups. This preliminary mosaic in true
X-ray colours is a fraction of the central deeper 2 sq.deg. area;
red: soft sources ($<2$ keV), blue: hard  sources ($>$ 2 keV). The
individual images have a diameter of  25  arcmin  and the exposure
time is 20 ks  on each field. The source density is found to be
$\sim$ 600 / sq.deg. in the [0.5-2] keV band (Valtchanov  et  al
2002, in preparation). It is interesting to note that this field
is devoid of sources in the ROSAT All-Sky Survey.}
\label{mpierre-wb1_fig5}
\end{figure*}
\begin{figure*}[h]
\centerline{\psfig{file=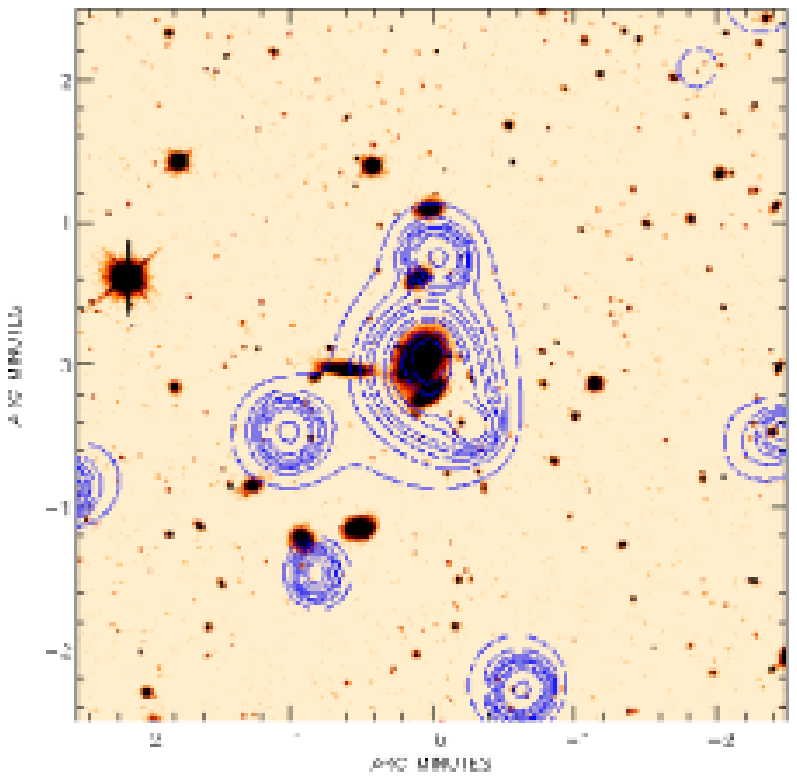,width=9cm,angle=0}
\psfig{file=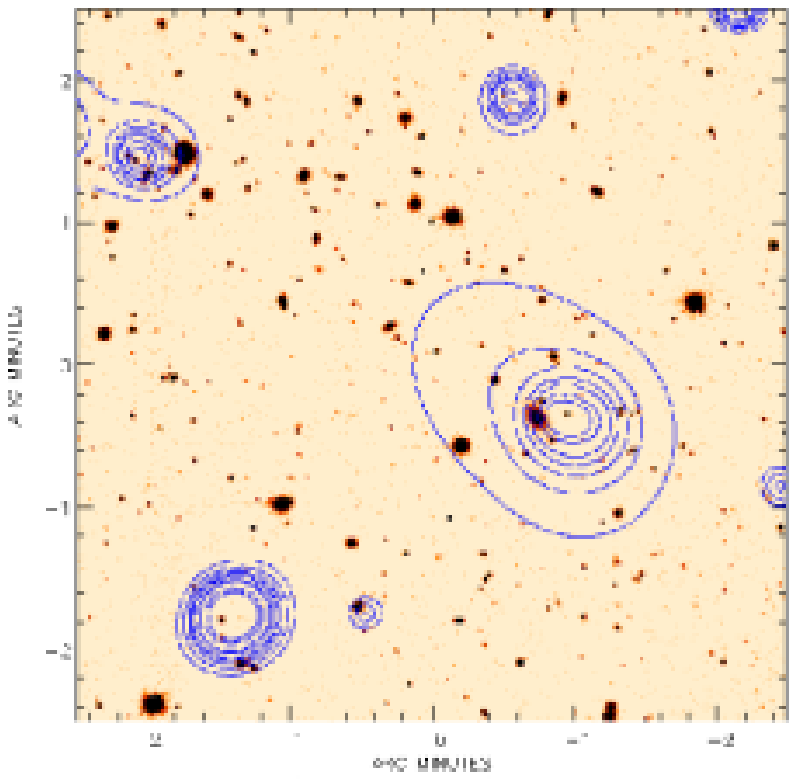,width=9cm,angle=0} }
\caption[]{Overlaid on a deep CFH12K I image (VIRMOS Deep Survey,
by Le F\`evre et al.), the XMM contours are shown in the [0.5-2]
keV band (after wavelet filtering).
 {\bf Left}: The central source is very soft
and found to be associated with a compact group in the optical.
This object was not not known prior to our observation, although
already flagged as a 2MASS source. Several other pointlike sources
are conspicuous on the image, but few of them have an obvious
optical identification. {\bf Right}: The clearly extended source
toward the image center has no obvious optical counterpart and,
therefore, is an interesting distant cluster candidate. Such a
situation enlightens the need for deep NIR photometry in order to
confirm the existence of $z > 1$ clusters, to provide photometric
redshifts and, eventually, galaxy targets for a spectroscopic
measurement} \label{mpierre-wb1_fig6}
\end{figure*}
\section{More information}
More information is available on the XMM-LSS web page:\\
 {\tt
http://vela.astro.ulg.ac.be/themes/spatial/xmm/\\
LSS/index\_e.html}

\end{document}